\documentclass[twocolumn,showpacs,preprintnumbers,amsmath,amssymb]{revtex4}

\usepackage{graphicx}

\begin{document}
% You should use BibTeX and revtex.bst for references
\bibliographystyle{revtex}
% marks overfull lines with blackboxes
\draft

%\preprint{}

\title{Single-crystal growth and dependences on the hole concentration and magnetic field of the magnetic ground state in the edge-sharing CuO$_2$ chain system Ca$_{2+x}$Y$_{2-x}$Cu$_5$O$_{10}$}
% Optional argument for running titles on pages
%\title[]{}

\author{K. Kudo}
\thanks{Present address: Institute for Materials Research, Tohoku University, Katahira 2-1-1, Aoba-ku, Sendai 980-8577, Japan}
\email{kudo@imr.tohoku.ac.jp}
\author{S. Kurogi}
\author{Y. Koike}
\email{koike@teion.apph.tohoku.ac.jp}
\affiliation{Department of Applied Physics, Graduate School of Engineering, Tohoku University, Aoba-yama 6-6-05, Aoba-ku, Sendai 980-8579, Japan}
\author{T. Nishizaki}
\author{N. Kobayashi}
\affiliation{Institute for Materials Research, Tohoku University, Katahira 2-1-1, Aoba-ku, Sendai 980-8577, Japan}

\date{\today}

\begin{abstract}
We have succeeded in growing large-size single-crystals of Ca$_{2+x}$Y$_{2-x}$Cu$_5$O$_{10}$ with $0 \le x \le 1.67$ and measured the magnetic susceptibility, specific heat and magnetization curve, in order to study the magnetic ground state in the edge-sharing CuO$_2$ chain as a function of hole concentration and magnetic field. 
In $0 \le x \le 1.3$, it has been found that an antiferromagnetically ordered phase with the magnetic easy axis along the $b$-axis is stabilized and that a spin-flop transition occurs by the application of magnetic fields parallel to the $b$-axis. 
The antiferromagnetic transition temperature decreases with increasing $x$ and disappears around $x =$ 1.4. 
Alternatively, a spin-glass phase appears around $x = 1.5$. 
At $x = 1.67$ where the hole concentration is $\sim$ 1/3 per Cu, it appears that a spin-gap state is formed owing to the formation of spin-singlet pairs. 
No sign of the coexistence of an antiferromagnetically ordered state and a spin-gap one suggested in Ca$_{1-x}$CuO$_2$ has been found in Ca$_{2+x}$Y$_{2-x}$Cu$_5$O$_{10}$. 
\end{abstract}

\pacs{75.40.-s, 75.40.Cx, 75.10.Jm}

\maketitle

\section{Introduction} 
Recently, it has been no more doubtful that the mechanism of the high temperature superconductivity links to the magnetism in the two-dimensional CuO$_2$ plane. 
Enormous volumes of experimental and theoretical works on the magnetism have revealed a variety of Cu$^{2+}$-spin states as a function of carrier doping\cite{Kastner}. 
The carrier doping drastically changes the antiferromagnetically ordered phase to a spin-glass phase, a stripe ordered one of spins and/or holes, a spin-gap one or a superconducting one. 
Therefore, the change of the magnetism through the carrier doping has attracted great interest.

\begin{figure}[t]
\begin{center}
\includegraphics[width=1\linewidth]{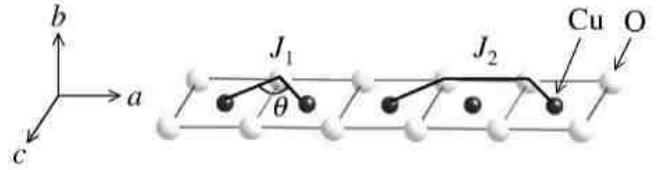}
\caption{Structure of the edge-sharing CuO$_2$ chain. The $\theta$ is the Cu-O-Cu bond angle. The $J_1$ and $J_2$ are the nearest and second nearest neighbor interactions between Cu$^{2+}$ spins, respectively. }
\label{fig:1}
\end{center}
\end{figure}
One-dimensional chain systems composed of Cu$^{2+}$ and O$^{2-}$ ions are carrier-dopable as well. 
They are categorized into corner-sharing chain systems or edge-sharing chain ones. 
In the former, CuO$_4$ squares are connected with each other by sharing a oxygen at the corner, while they are connected by sharing two oxygens at the edge in the latter. 
Doped holes make CuO$_4$ squares nonmagnetic through the change of the Cu valency from Cu$^{2+}$ to Cu$^{3+}$ or the formation of Zhang-Rice singlet pairs of oxygen holes and Cu$^{2+}$ spins\cite{Zhang}, to divide the magnetic intrachain interaction. 
The layered cuprate Sr$_{14}$Cu$_{24}$O$_{41}$\cite{Dagotto} possesses pre-doped edge-sharing CuO$_2$ chains\cite{Osafune} whose spin-state exhibits a spin gap\cite{Tsuji,KumagaiPRL,Takigawa,TsujiJLTP,Eccpoly,Ecc,Matsuda14n,Kato,Kato2,Kato3,Matsuda14x,Carter}. 
With decreasing hole-concentration in the chain through the substitution of La$^{3+}$, Y$^{3+}$ or Ca$^{2+}$ for Sr$^{2+}$, the Cu$^{2+}$-spin state changes from a spin-gap state to an antiferromagnetically ordered one\cite{TsujiJLTP, Matsuda14n,MHiroi,MIsobe,Nagata}. 
However, the study of the carrier-doping effect in the chain is not easy in Sr$_{14}$Cu$_{24}$O$_{41}$ because of the existence of the spin-ladder plane of Cu$_2$O$_3$. 
A simple structure system composed of carrier-dopable edge-sharing CuO$_2$ chains is Ca$_{1-x}$CuO$_2$, whose hole-concentration is controlled by changing $x$. 
According to the theoretical calculation\cite{Mizuno}, the magnetic interaction between the nearest neighbor Cu$^{2+}$ spins in the edge-sharing CuO$_2$ chain is antiferromagnetic for $\theta > 95^{\circ}$, while it is ferromagnetic for $\theta < 95^{\circ}$, where $\theta$ is the Cu-O-Cu bond angle as shown in Fig. \ref{fig:1}. 
The $\theta$ value in Ca$_{1-x}$CuO$_2$ has been estimated as $\sim$ 95$^\circ$ which is located at the boundary\cite{Babu,theta}, implying that an attractive spin-state may appear. 
Actually, a novel coexistence of an antiferromagnetically ordered state and a spin-gap one has been suggested in a range of $0.164 \leq x \leq 0.190$ from the magnetic susceptibility and specific heat measurements\cite{Hiroi} and also from the structural analysis\cite{Isobe}. 
In order to explain the coexistence, Hiroi {\it et al.}\cite{Hiroi} have proposed a two-sublattice model in which a single chain is divided into two independent chains, taking into account the theoretical result that the magnetic interaction between the second nearest neighbor Cu$^{2+}$ spins, $J_2$, is antiferromagnetic and that $|J_2|$ is much larger than the absolute value of the nearest neighbor interaction, $J_1$\cite{Mizuno}. 
That is, every other Cu$^{2+}$ spin belonging to one sublattice is regarded as forming an antiferromagnetic long-range order, while holes and Cu$^{2+}$ spins belonging to the other sublattice are regarded as being localized and forming spin-singlet pairs with a spin gap, respectively. 
From the structural analysis, on the other hand, Isobe {\it et al.}\cite{Isobe} have proposed another model in which holes are located almost periodically at intervals of two Cu$^{2+}$ spins. 
In this case, it follows that Cu$^{2+}$ spins on both sides of a hole form a spin-singlet pair with a spin gap, while a small amount of residual Cu$^{2+}$ spins exhibit an antiferromagnetic order. 
The interpretation on the coexistence of the two states in Ca$_{1-x}$CuO$_2$ has not yet been settled. 
However, detailed experiments have not been carried out, because it is very hard to grow single crystals of Ca$_{1-x}$CuO$_2$.

\begin{figure}[t]
\begin{center}
\includegraphics[width=0.7\linewidth]{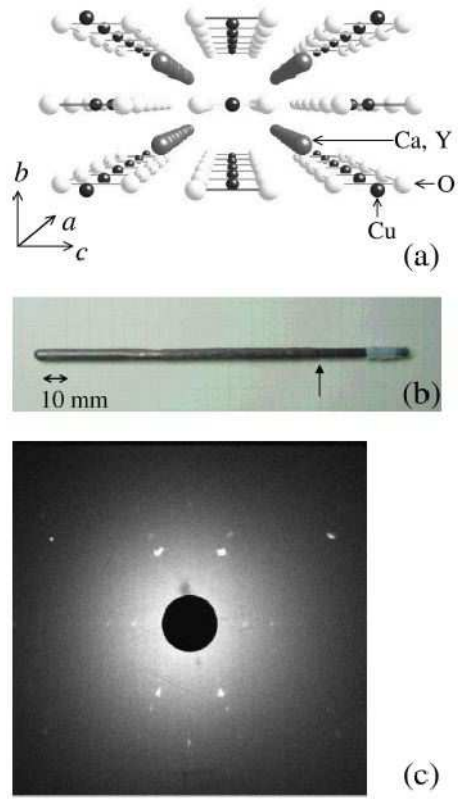}
\caption{
(a) Schematic crystal structure of Ca$_{2+x}$Y$_{2-x}$Cu$_5$O$_{10}$.
(b) Picture of a single crystal of Ca$_{2+x}$Y$_{2-x}$Cu$_5$O$_{10}$ with $x =$ 1.5 grown by the TSFZ method, which is a left part of the rod from the upward arrow. 
(c) X-ray back-Laue photography of a grown crystal of Ca$_{2+x}$Y$_{2-x}$Cu$_5$O$_{10}$ with $x =$ 1.5 in the x-ray parallel to the $c$-axis. }
\label{fig:2}
\end{center}
\end{figure}

A similar edge-sharing CuO$_2$ chain system is Ca$_{2+x}$Y$_{2-x}$Cu$_5$O$_{10}$. 
The $\theta$ value at $x = 0$ has been estimated as $\sim 91^\circ$\cite{Miyazaki}. 
With increasing $x$, $\theta$ increases and reaches $\sim 93.4^\circ$ at $x = 2$\cite{Miyazaki}, which is comparable with that of Ca$_{1-x}$CuO$_2$\cite{theta}. 
Therefore, a similar coexistence of an antiferromagnetically ordered state and a spin-gap one is expected in  Ca$_{2+x}$Y$_{2-x}$Cu$_5$O$_{10}$ with large $x$ values as well.  
As shown in Fig. \ref{fig:2}(a), it possesses a layered structure composed of stacked $ac$-planes where edge-sharing chains run along the $a$-axis. 
The magnetic properties have been studied so far mainly using polycrystalline samples. 
It has been found from the magnetic susceptibility\cite{Hayashi,Miyazaki,Matsudax}, specific heat\cite{Matsudax} and neutron scattering measurements\cite{Matsudan,Fong} that the sample without any hole carriers at $x =$ 0 exhibits an antiferromagnetic order at low temperatures below the antiferromagnetic transition temperature $T_{\rm N} =$ 30 K and that Cu$^{2+}$ spins are arranged ferromagnetically along the chain and the inter-chain coupling is antiferromagnetic. 
With increasing $x$, namely, with the increase of doped holes, $T_{\rm N}$ decreases and finally disappears around $x = 1.5$. 
The single-crystal growth of Ca$_{2+x}$Y$_{2-x}$Cu$_5$O$_{10}$ has been reported by Oka {\it et al.}\cite{Oka} for $x =$ 0 and 0.5. 
The magnetic susceptibility measurements of these crystals have revealed that the magnetic easy axis is the $b$-axis\cite{Oka,Yamaguchi}. 
Moreover, the magnetic dispersion relation has been clarified from the neutron scattering experiment\cite{Matsudans}. 
The single crystal growth for $x >$ 0.5 has also been reported\cite{Oka2}, but the grown crystals are too small to be available for the detailed measurements, though the novel ground state in Ca$_{1-x}$CuO$_2$ is expected to appear at $x >$ 0.5 in Ca$_{2+x}$Y$_{2-x}$Cu$_5$O$_{10}$. 

Therefore, we have tried to grow  large-size single-crystals of Ca$_{2+x}$Y$_{2-x}$Cu$_5$O$_{10}$ with $x > 0.5$  and succeeded in the growth up to $x =$ 1.67.
In this paper, we report results of the single-crystal growth and the magnetic susceptibility, specific heat and magnetization curve measurements using the single crystals of $x =$ 0, 0.5, 1.0, 1.2, 1.3, 1.5 and 1.67, and discuss the hole-concentration and magnetic-field dependences of the magnetic ground state of Ca$_{2+x}$Y$_{2-x}$Cu$_5$O$_{10}$. 
The preliminary results have already been reported by Kurogi {\it et al.}\cite{Kurogi}.

\begin{table*}[htb]
\caption{Growth conditions, chemical compositions and dimensions of Ca$_{2+x}$Y$_{2-x}$Cu$_5$O$_{10}$ single crystals.}
\begin{ruledtabular}
\begin{tabular}{ccccccc}
               & nominal composition & solvent                      & growth rate  &                                      & composition (ICP-AES)              &  dimensions                                \\ 
$x$       & Ca : Y : Cu                      & Ca : Y : Cu                 &    ( mm/h )   &  atmosphere              & Ca : Y : Cu                  &  ( mm$\phi \times$ mm )     \\ 
\hline
0           & 2.0 : 2.0 : 5.0                  & 1.00 : 0.45 : 5.10   & 0.50               & air 1 atm                     & 1.94 : 2.09 : 4.96  & 6 $\times$ 60                         \\
0.5       & 2.5 : 1.5 : 5.0                  & 1.00 : 0.15 : 4.20   & 0.50               & O$_2$ 1 atm            & 2.51 : 1.52 : 4.97  & 6 $\times$ 60                         \\
1.0       & 3.0 : 1.0 : 5.0                  & 1.00 : 0.10 : 3.50   & 0.40               & O$_2$ 6 atm            & 2.99 : 1.05 : 4.96  & 6 $\times$ 60                         \\
1.2       & 3.2 : 0.8 : 5.0                  & 1.00 : 0.10 : 4.02   & 0.35              & O$_2$ 10 atm         & 3.21 : 0.78 : 5.01  & 6 $\times$ 40                         \\
1.3       & 3.3 : 0.7 : 5.0                  & 1.00 : 0.10 : 4.28   & 0.35              & O$_2$ 10 atm         & 3.29 : 0.69 : 5.02  & 6 $\times$ 40                         \\
1.5       & 3.5 : 0.5 : 5.0                  & 1.00 : 0.10 : 4.80   & 0.35              & O$_2$ 10 atm         & 3.52 : 0.55 : 4.93  & 6 $\times$ 40                         \\
1.67   & 3.667 : 0.333 : 5.0        & 1.00 : 0.07 : 4.08    & 0.35              & O$_2$ 10 atm         & 3.64 : 0.32 : 5.04  & 6 $\times$ 40                         \\ 
\end{tabular}
\end{ruledtabular}
\label{tbl:1}
\end{table*}
\section{experimental}
Single crystals of Ca$_{2+x}$Y$_{2-x}$Cu$_5$O$_{10}$ were grown by the Traveling-Solvent Floating-Zone (TSFZ) method.
In order to prepare the feed rod for the TSFZ growth, first, we prepared the polycrystalline powder of Ca$_{2+x}$Y$_{2-x}$Cu$_5$O$_{10}$ by the solid-state reaction method.
The prescribed amount of CaCO$_3$, Y$_2$O$_3$ and CuO powders with 99.9 \% purity was mixed, ground and prefired at 900 $^\circ$C in air for 12 h.
After pulverization, the prefired powder was mixed and sintered at 1000 $^\circ$C for 1 week with several times of intermediate grinding.
After 1 h grinding, the powder was isostatically cold-pressed at 400 bar into a rod of 7 mm in diameter and 150 mm in length. 
Then, the rod was sintered at 1000 $^\circ$C in air for 1 day. 
As a result, a tightly and densely sintered feed rod was prepared. 
As Ca$_{2+x}$Y$_{2-x}$Cu$_5$O$_{10}$ melts incongruently\cite{Oka,Oka2}, solvent disks with different compositions were prepared sintering at 900 $^\circ$C in air for 12 h. 
The composition of the solvent listed in Table \ref{tbl:1} was determined referring to the previous report by Oka {\it et al.}\cite{Oka2}. 
The TSFZ growth was carried out with the obtained feed rod and a disk of the solvent material in an infrared heating furnace equipped with a quartet ellipsoidal mirror (Crystal Systems Inc., Model FZ-T-4000-H). 
The rotation speed of the upper and lower shafts was 10 rpm in the opposite direction to secure the homogeneity of the liquid in the molten zone. 
The rotation of the lower shaft was stopped when the $ac$-plane became visible during the growth\cite{stop}. 
The growth rate and the atmosphere listed in Table \ref{tbl:1} were optimized for each $x$ by trial and error. 
The grown crystals were then characterized using the x-ray back-Laue photography and were confirmed having a single phase by the powder x-ray diffraction. 
The chemical compositions were determined by the inductively coupled plasma atomic emission spectroscopy (ICP-AES).

The temperature dependence of the magnetic susceptibility was measured in a magnetic field of 1 T, using a SQUID magnetometer (Quantum Design, Model MPMS).
The specific heat measurements were carried out in magnetic fields up to 9 T by the thermal relaxation technique (Quantum Design, Model PPMS). 
The magnetization curve in magnetic fields up to 14 T was measured using a vibrating sample magnetometer (Oxford Instruments, Model MagLab).

\section{Results and Discussion}
\subsection{Single-crystal growth}
\begin{figure}[h]
\begin{center}
\includegraphics[width=0.5\linewidth]{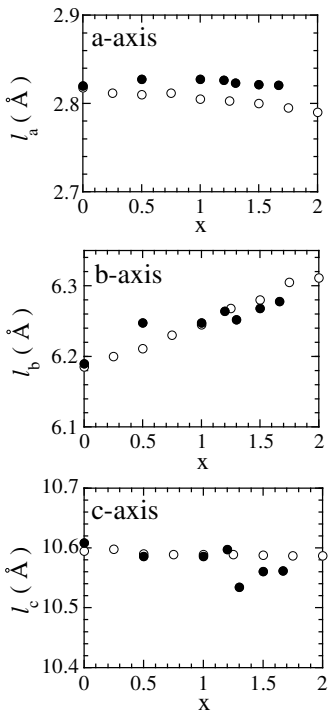}
\caption{Hole-concentration dependence of the lattice constants, $l_{\rm a}$, $l_{\rm b}$ and $l_{\rm c}$, parallel to the $a$-, $b$- and $c$-axes of Ca$_{2+x}$Y$_{2-x}$Cu$_5$O$_{10}$, respectively. Closed and open circles indicate data of the present work and Hayashi {\it et al.}\cite{Hayashi}, respectively. }
\label{fig:3}
\end{center}
\end{figure}

Figure \ref{fig:2}(b) shows a picture of a grown single-crystal of Ca$_{2+x}$Y$_{2-x}$Cu$_5$O$_{10}$ with $x =$ 1.5. 
The cross section is ellipsoidal in shape, owing to the layered structure stacking along the $b$-axis. 
Figure \ref{fig:2}(c) shows the x-ray back-Laue photography in the x-ray parallel to the $c$-axis. 
The diffraction spots exhibit a two-fold symmetry and are very sharp, indicating the good quality of the single crystal. 
The typical dimensions of a single domain are 6 mm$\phi$ $\times$ (40--60) mm, as listed in Table \ref{tbl:1} for various $x$ values. 
The dimensions of single crystals with $1.0 \le x \le 1.67$ are as large as those with $x =$ 0 and 0.5. 
The powder x-ray diffraction patterns of the single crystals with $0 \le x \le 1.67$ reveal no impurity phases. 
Figure \ref{fig:3} shows the $x$ dependence of the lattice constants $l_{\rm a}$, $l_{\rm b}$ and $l_{\rm c}$ parallel to the $a$- and $b$- and $c$-axes, respectively. 
With increasing $x$, $l_{\rm b}$ tends to increase, while both $l_{\rm a}$ and $l_{\rm c}$ tend to decrease slightly. 
Taking into account that the substitution of Ca$^{2+}$ (ionic radius = 0.99 {\rm \AA}) for Y$^{3+}$ (0.92 {\rm \AA}) expands the distance between $ac$-planes and that induced holes in the CuO$_2$ chain shorten the distance between Cu$^{2+}$ and O$^{2-}$ ions in the $ac$-plane, the substitution is regarded as being successful. 
The $x$ dependence is in rough agreement with that of polycrystalline samples\cite{Hayashi}. 
As listed in Table \ref{tbl:1}, the chemical composition determined by ICP-AES almost coincides with the nominal composition also. 
Thus, it is said that the growth of large-size single crystals of Ca$_{2+x}$Y$_{2-x}$Cu$_5$O$_{10}$ with $0 \le x \le 1.67$ is successful.

Here, we note the reasons why we have succeeded in growing single crystals of Ca$_{2+x}$Y$_{2-x}$Cu$_5$O$_{10}$ with large $x$ values by the TSFZ method.
The first is that the high oxygen pressure of 10 atm was applied. 
Oka {\it et al.}\cite{Oka2} have performed the TSFZ growth under the oxygen pressure up to 6 atm and reported that the oxygen pressure tends to suppress the formation of impurity phases. 
The high oxygen pressure of 10 atm might suppress the formation of impurity phases in our trial as well. 
The second is that a seed of a single crystal was used. 
The third is that the rotation of the lower rod was stopped when the $ac$-plane appeared. 
Although the effect has not been clarified, it is a matter of fact that the growth was not successful for $x >$ 1.0 without this procedure\cite{stop}. 
The fourth is that the growth rate was rather small. 
In fact, there existed many cracks in the single crystals grown at a rapid rate, which may be due to the small supersaturation on account of the narrow liquidus line.

\subsection{Hole-concentration dependence of the magnetic ground state}
\begin{figure*}[tp]
\begin{center}
\includegraphics[width=1\linewidth]{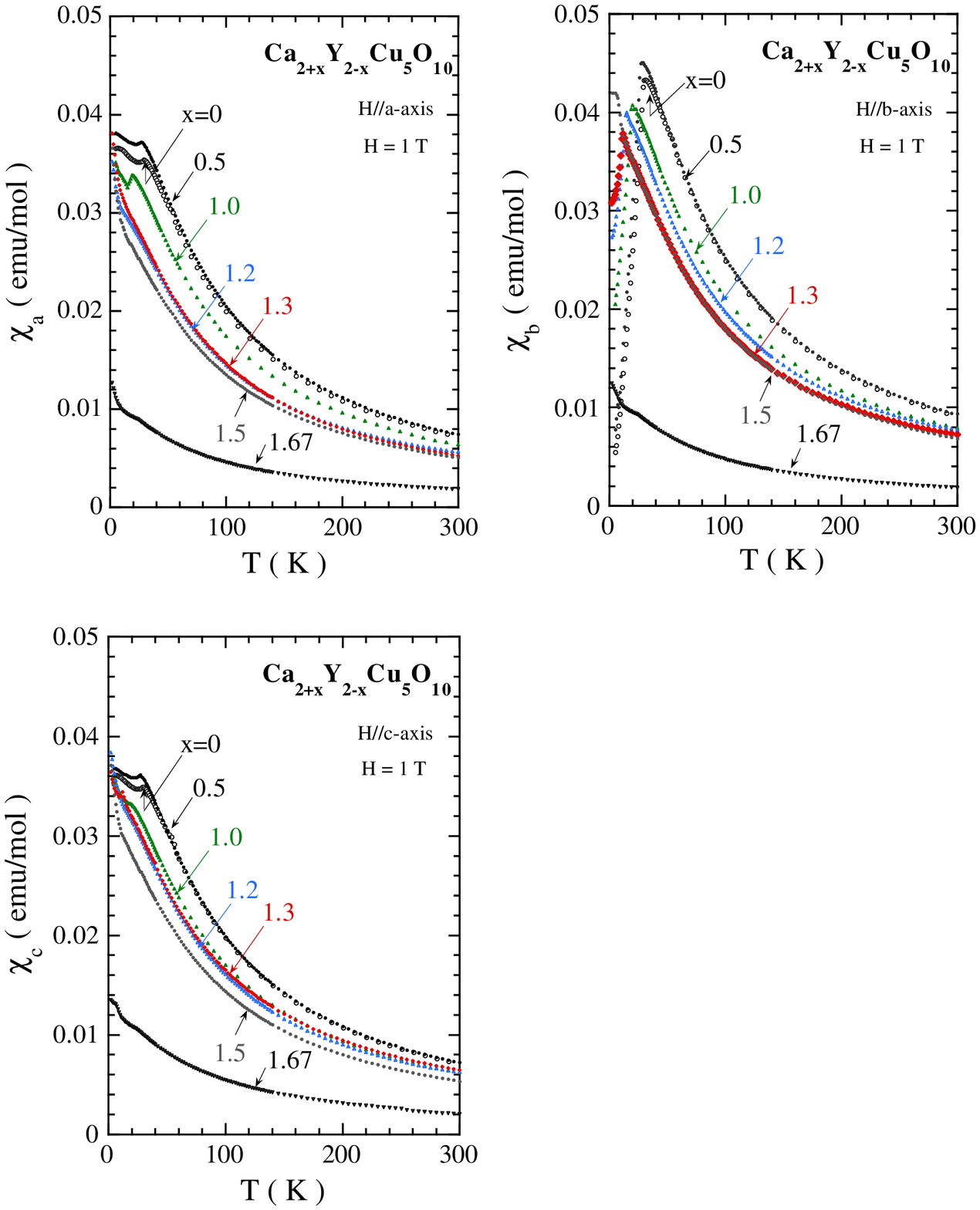}
\caption{(Color online) Temperature dependence of the magnetic susceptibilities, $\chi_{\rm a}$, $\chi_{\rm b}$ and $\chi_{\rm c}$, of  Ca$_{2+x}$Y$_{2-x}$Cu$_5$O$_{10}$ with $0 \le x \le 1.67$ in a magnetic field of 1 T parallel to the $a$-, $b$- and $c$-axes, respectively.}
\label{fig:4}
\end{center}
\end{figure*}
\begin{figure*}[tp]
\begin{center}
\includegraphics[width=0.9\linewidth]{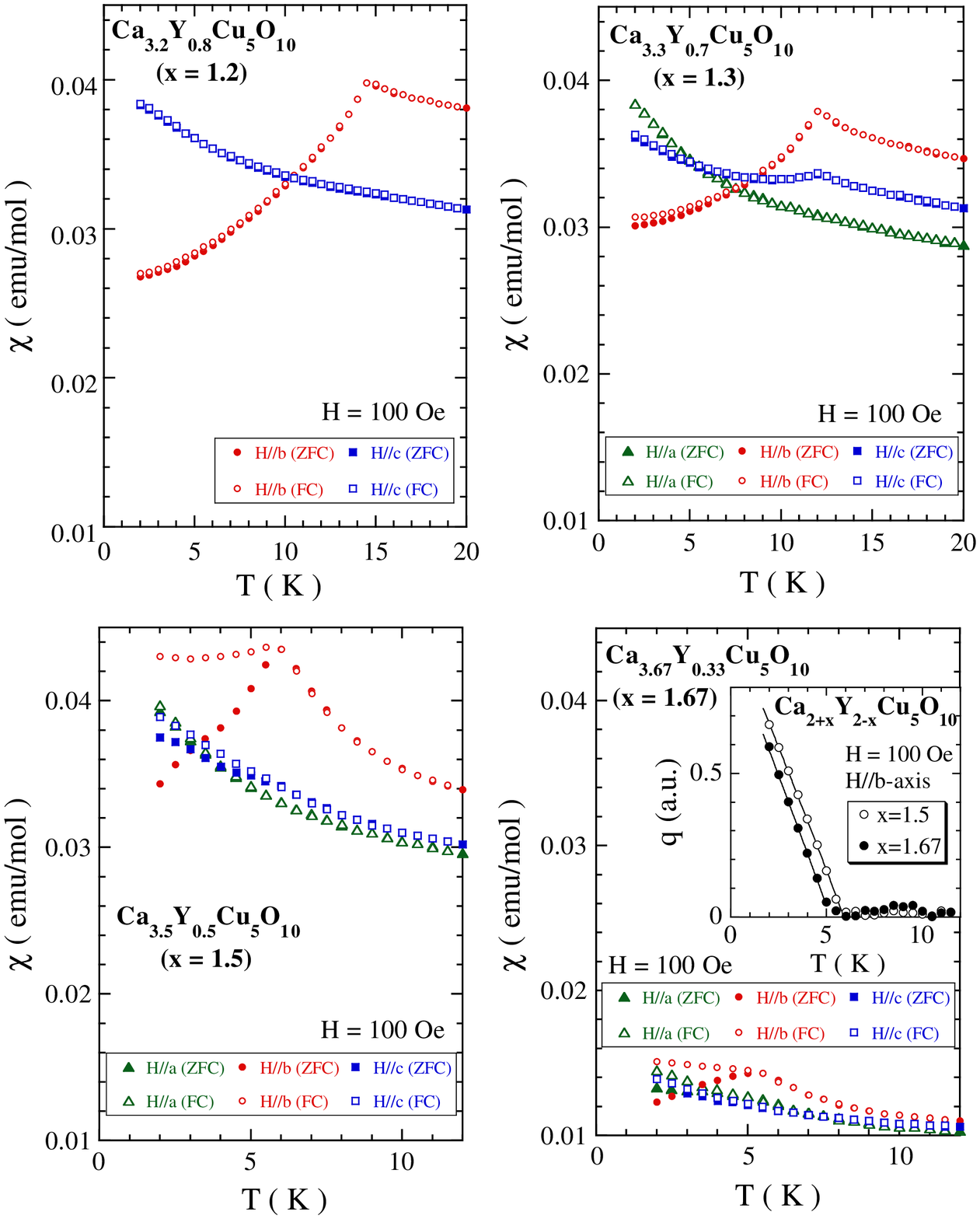}
\caption{(Color online)Temperature dependence of the magnetic susceptibilities, $\chi_{\rm a}$, $\chi_{\rm b}$ and $\chi_{\rm c}$, of  Ca$_{2+x}$Y$_{2-x}$Cu$_5$O$_{10}$ with $1.2 \le x \le 1.67$ on both zero-field cooling and field cooling in a magnetic field of 100 Oe. Closed and open symbols correspond to the data on zero-field cooling and field-cooling, respectively. The inset shows the temperature dependence of the spin-glass order parameter, $q$, in $x = 1.5$ and 1.67. 
Solid lines denote fitting curves proportional to $(T_{\rm spin-glass} - T)^\beta$.}
\label{fig:5}
\end{center}
\end{figure*}
Figure \ref{fig:4} shows the temperature dependence of the magnetic susceptibilities, $\chi_{\rm a}$, $\chi_{\rm b}$ and $\chi_{\rm c}$, of Ca$_{2+x}$Y$_{2-x}$Cu$_5$O$_{10}$ with $0 \le x \le 1.67$ in a magnetic field of 1 T parallel to the $a$-, $b$- and $c$-axes, respectively. 
In $x =$ 0, 0.5, 1.0, 1.2 and 1.3, it is found that $\chi_{\rm b}$ increases with decreasing temperature and exhibits a sharp peak indicating the antiferromagnetic transition. 
The $T_{\rm N}$ is determined at the peak to be 31 K and 29 K for $x =$ 0 and 0.5, respectively, which is similar to the result by Yamaguchi {\it et al.}\cite{Yamaguchi}.
Furthermore, $T_{\rm N}$ is estimated as 20 K, 15 K and 12 K for $x =$ 1.0, 1.2 and 1.3, respectively. 
Both $\chi_{\rm a}$ and $\chi_{\rm c}$ tend to become constant below $T_{\rm N}$, so that the $b$-axis is regarded as the magnetic easy axis in $0 \le x \le 1.3$. 
In $x = 1.5$ and 1.67, on the other hand, no antiferromagnetic transition is observed, but a small and broad shoulder appears around 20 K not only in $\chi_{\rm b}$ but also in $\chi_{\rm a}$ and $\chi_{\rm c}$. 
The isotropic broad shoulder is not characteristic of the antiferromagnetic long-range order but analogous to a broad peak observed in a spin-gap state\cite{Kato,Kato2,Kato3,Matsuda14x,Carter,Hiroi,Isobe}. 
Here, note that the value of the magnetic susceptibility in $x = 1.67$ is remarkably small, which is discussed later.

Figure \ref{fig:5} shows the temperature dependence of $\chi_{\rm a}$, $\chi_{\rm b}$ and $\chi_{\rm c}$ on both zero-field cooling and field cooling in a magnetic field of 100 Oe. 
The temperature dependence exhibits a hysteresis in $1.3 \le x \le 1.67$, while it is reversible in $x =$ 1.2. 
The hysteresis indicates a spin-glass transition with the transition temperature $T_{\rm spin-glass} \sim$ 6 K.
$T_{\rm spin-glass}$ is independent of $x$. 
Moreover, the hysteresis is much smaller in $x = 1.3$ and 1.67 than in $x = 1.5$. 
Accordingly, the spin-glass phase observed in $x = 1.3$ and 1.67 may not be a main phase but a minor one due to the inhomogeneity of $x$ in a crystal. 
As shown in the inset of Fig. \ref{fig:5}, in fact, the temperature dependence of the spin-glass order parameter, $q$, of $x = 1.67$ defined by $\chi = \chi_0 + (1 - q)C/T$\cite{Chou,Wakimoto} almost coincides with that of $x = 1.5$, indicating that the spin-glass phase of $x = 1.67$ quite resembles that of $x = 1.5$. 
Here, $\chi_0$ and $C$ are the constant susceptibility and the Curie constant, respectively.
From the fitting analysis using $q \sim (T_{\rm spin-glass} - T)^\beta$, the critical exponent $\beta$ is estimated as 0.97 and 0.98 for $x =$ 1.5 and 1.67, respectively. 
These $\beta$ values are close to the mean field prediction $\beta = 1$ and to experimental values, such as 0.7 and 0.9 for Cu:Mn\cite{Omari} and 0.9--0.97 for La$_{2-x}$Sr$_x$CuO$_4$ with $0.03 \le x \le 0.05$\cite{Chou,Wakimoto}.

\begin{figure}[htb]
\begin{center}
\includegraphics[width=1\linewidth]{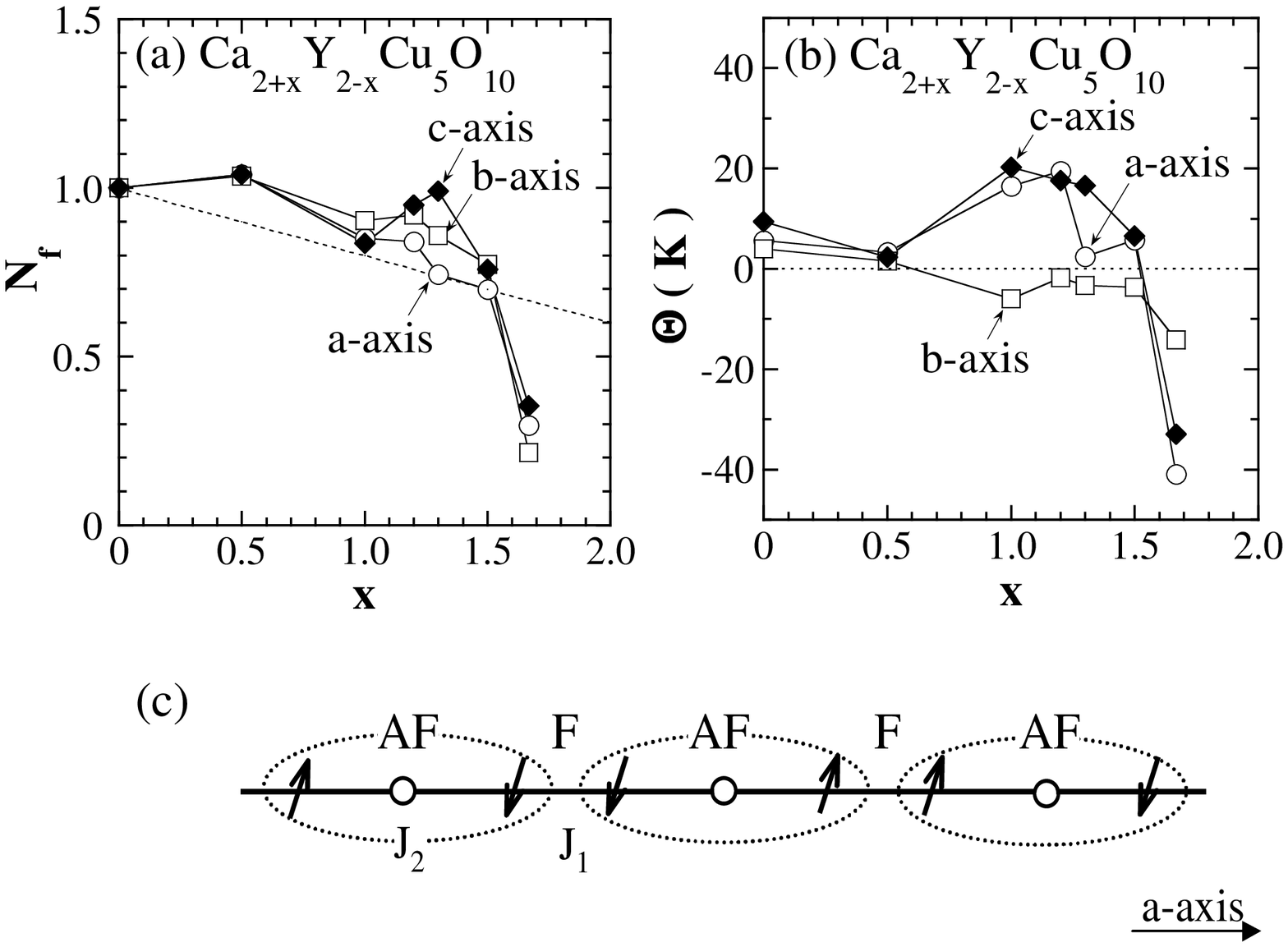}
\caption{Dependences on $x$ of (a) the number of free spins per Cu, $N_{\rm f}$, and (b) the Weiss temperature, $\Theta$, in  Ca$_{2+x}$Y$_{2-x}$Cu$_5$O$_{10}$. 
(c) Possible picture of spin-singlet pairs in $x = 1.67$. Arrows and circles indicate Cu$^{2+}$ spins and holes, respectively. 
The dotted line in (a) indicates $N_{\rm f}$ estimated from the chemical formula. 
Dotted ellipses in (c) indicate spin-singlet pairs. 
}
\label{fig:6}
\end{center}
\end{figure}
Here, we discuss small values of both $\chi_{\rm a}$ and $\chi_{\rm b}$ and $\chi_{\rm c}$ in $x = 1.67$. 
For all the measured single-crystals, the temperature dependence of the magnetic susceptibility, $\chi$, at high temperatures between 250 K and 300 K is well fitted using the Curie-Weiss law, $\chi = \chi_0 + N_{\rm f}Ng^2\mu_{\rm B}^2S(S + 1)/[3k_{\rm B}(T - \Theta)]$, where $N_{\rm f}$ is the number of free spins per Cu, $N$ the number of Cu atoms, $g$ the $g$-factor, $\mu_{\rm B}$ the Bohr magneton, $S$ the spin quantum number, $k_{\rm B}$ the Boltzmann constant and $\Theta$ the Weiss temperature. 
The $g$ value is estimated from the fitting of the data of $x = 0$ where $N_{\rm f} = 1$. 
It is dependent on the field direction and estimated as $g_{\rm a} = 2.13 (H$$\parallel$$a$-axis), $g_{\rm b} = 2.39 (H$$\parallel$$b$-axis) and $g_{\rm c} = 2.10 (H$$\parallel$$c$-axis). 
Using these values, values of $N_{\rm f}$ and $\Theta$ are estimated, as plotted in Figs. \ref{fig:6}(a) and (b). 
It is found that the obtained value of $N_{\rm f}$ is comparable with that estimated from the chemical formula for $0 \le x \le$ 1.5, while $N_{\rm f}$ is roughly one half or one third of the latter for $x = 1.67$. 
The $\Theta$ value in $x = 1.67$ is $-30 \pm 10$ K, whose absolute value is remarkably large compared with those of the other $x$ values. 
The marked decrease of $N_{\rm f}$ in $x = 1.67$ may be explained as follows. 
According to the theoretical result\cite{Mizuno}, magnetic interactions between Cu$^{2+}$ spins, expressed as $J_1$ and $J_2$ in Fig. \ref{fig:1} and Fig. \ref{fig:6}(c), are expected to be ferromagnetic and antiferromagnetic, respectively, as mentioned in Sec. I.  
Therefore, this spin system is regarded as being highly frustrated. 
As for $x = 1.67$, the hole concentration is $\sim$ 1/3 per Cu. 
Therefore, when holes are located in order at every third site in the chain, it follows that all the Cu$^{2+}$ spins can form such spin-singlet dimers as shown in Fig. \ref{fig:6}(c) without residual spins and without the frustration between $J_1$ and $J_2$, leading to the marked decrease of $N_{\rm f}$\cite{residual}. 
As mentioned in Sec. I, a similar arrangement of holes and spins has been proposed by Isobe {\it et al.}\cite{Isobe} from the structural analysis of Ca$_{1-x}$CuO$_{2+\delta}$ with $x = 0.176$, whose hole-concentration is close to that of Ca$_{2+x}$Y$_{2-x}$Cu$_5$O$_{10}$ with $x =$ 1.67. 
Accordingly, the magnetic ground state in $x = 1.67$ may be spin-singlet pairs with a spin gap. 
To be conclusive, the neutron scattering experiment for $x = 1.67$ is under way\cite{Matsudasingle}.

\begin{figure}[t]
\begin{center}
\includegraphics[width=1\linewidth]{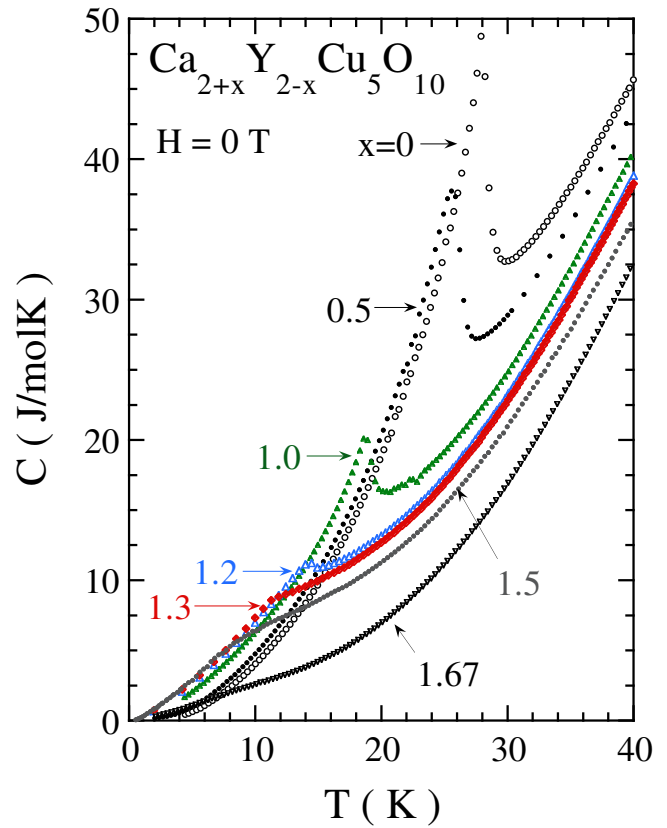}
\caption{(Color online) Temperature dependence of the specific heat, $C$,  of  Ca$_{2+x}$Y$_{2-x}$Cu$_5$O$_{10}$ with $0 \le x \le 1.67$ in zero field. 
}
\label{fig:7}
\end{center}
\end{figure}
Figure \ref{fig:7} shows the temperature dependence of the specific heat, $C$, of Ca$_{2+x}$Y$_{2-x}$Cu$_5$O$_{10}$ in zero field. 
A $\lambda$-type peak is observed in $x =$ 0, 0.5, 1.0, 1.2 and 1.3, corresponding to the antiferromagnetic transition.  
The $T_{\rm N}$ is estimated at the peak to be 29 K, 26 K, 18 K, 15 K and 12 K for $x =$ 0, 0,5, 1.0, 1.2 and 1.3, respectively. 
These values are consistent with those estimated from the magnetic susceptibility measurements. 
The $\lambda$-type peak is smeared and shifted to lower temperatures with increasing $x$, and finally disappears at $x =$ 1.5. 
Formerly, Chabot {\it et al.}\cite{Chabot} have stated from the specific heat measurements of polycrystalline samples that the temperature dependence of the specific heat in $x =$ 1.0 is regarded as a behavior characteristic of the one-dimensional Heisenberg antiferromagnetic chain, taking into account the result that no $\lambda$-type peak is observed at $T \geq$ 0.5 K. 
The difference may be attributed to the quality of the samples, because it is described in Ref. \onlinecite{Chabot} that a large Curie tail has been observed in the temperature of the magnetic susceptibility at $x > 0$, while no Curie tail is observed in our samples. 
As for $x =$ 1.5 and 1.67, the specific heat exhibits a broad peak around 10 K instead of a $\lambda$-type peak.  
Such a peak may be regarded as being due to a static short-range order like a spin glass or due to a spin gap caused by the formation of spin-singlet pairs. 
Taking into account the result that the onset temperature of the broad peak of the specific heat ($\sim$ 20 K) coincides with the temperature of the broad shoulder observed in the temperature dependence of the magnetic susceptibility, the origin seems to be the spin-gap formation. 
However, this is inconsistent with the result that the broad peak in $x = 1.67$ is smaller than that in $x = 1.5$, because the number of spin-singlet pairs in $x = 1.67$ should be larger than that in $x = 1.5$. 
Taking into account the results that the spin-glass phase in $x = 1.67$ may be a minor phase and that the broad peak in $x = 1.5$ appears to systematically change from the $\lambda$-type peak of $x \le 1.3$, the origin is probably the formation of a static short-range order like a spin glass.

Here, we would like to calculate the entropy of Cu$^{2+}$ spins, $S_{\rm spin}$, from the specific heat of spins, $C_{\rm spin}$, as $S_{\rm spin} = \int_0^T C_{\rm spin}/T' \ {\rm d}T'$. 
However, the precise evaluation of $S_{\rm spin}$ is hard, because the specific heat of phonons, $C_{\rm phonon}$, is unknown. 
Here, suffice it to say that $S_{\rm spin}$ is likely to exhibit the minimum at $x = 1.67$ among the present samples, assuming that $C_{\rm phonon}$ is independent of $x$. 
This is qualitatively consistent with the small value of $N_{\rm f}$ in $x = 1.67$ shown in Fig. \ref{fig:6}(a). 
Therefore, it is concluded that the specific heat data also indicate the formation of spin-singlet pairs with a spin gap at $x = 1.67$.

\begin{figure}[t]
\begin{center}
\includegraphics[width=1\linewidth]{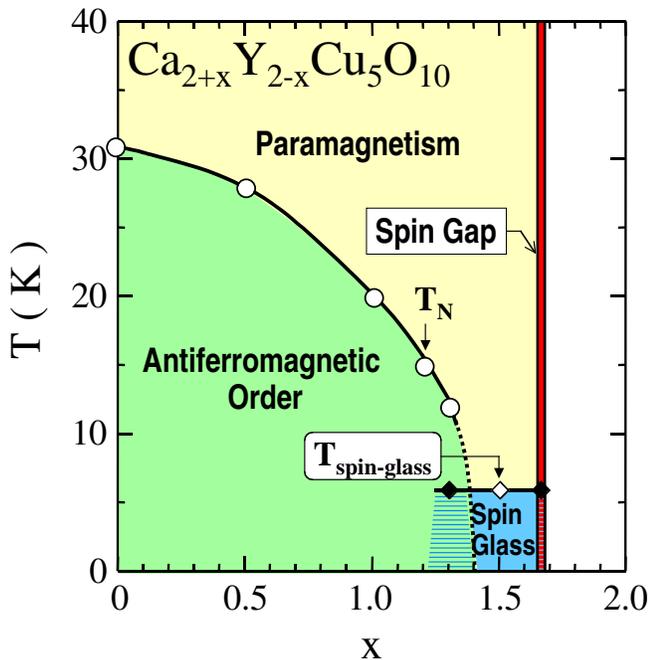}
\caption{(Color online) Magnetic phase diagram of Ca$_{2+x}$Y$_{2-x}$Cu$_5$O$_{10}$ in zero field. 
Open and closed symbols indicate temperatures below which the main phase and the minor phase appear, respectively.}
\label{fig:8}
\end{center}
\end{figure}
Figure \ref{fig:8} summarizes the $x$ dependence of the characteristic temperatures in zero field for Ca$_{2+x}$Y$_{2-x}$Cu$_5$O$_{10}$. 
An antiferromagnetically ordered phase is found in small $x$ values. 
The $T_{\rm N}$ decreases with increasing $x$ and disappears around 1.4. 
Alternatively, a spin-glass phase appears around $x = 1.5$. 
The spin-glass phase observed in $x = 1.3$ and 1.67 may be a minor phase. 
In $x =$ 1.67, it appears that more than half of Cu$^{2+}$ spins form spin-singlet pairs with a spin gap in the measured temperature-range.

Referring to Ref. \onlinecite{Miyazaki}, the $\theta$ value of Ca$_{2+x}$Y$_{2-x}$Cu$_5$O$_{10}$ with $x = 1.67$ is estimated as $\sim93^\circ$, which is comparable with that of Ca$_{1-x}$CuO$_2$\cite{theta}. 
Therefore, not only hole concentrations but also $\theta$ values in these systems are almost similar to each other. 
However, no sign of the coexistence of an antiferromagnetically ordered state and a spin gap one suggested in Ca$_{1-x}$CuO$_2$ has been found in Ca$_{2+x}$Y$_{2-x}$Cu$_5$O$_{10}$. 
This may be because the intense modulation in the $ac$-plane and/or the randomness of Ca$^{2+}$ and Y$^{3+}$ ions in Ca$_{2+x}$Y$_{2-x}$Cu$_5$O$_{10}$ suppress the mobility of holes so as not to make any suitable arrangement of holes for the formation of the coexistence state. 
In fact, the modulation periods parallel to the $a$- and $c$-axes in Ca$_{2+x}$Y$_{2-x}$Cu$_5$O$_{10}$ with $x = 1.67$ are as small as $\sim$ 90 \% and $\sim$ 40 \% of those in Ca$_{1-x}$CuO$_2$, respectively\cite{Miyazaki}.

\subsection{Magnetic-field dependence of the magnetic ground state}
\begin{figure*}[t]
\begin{center}
\includegraphics[width=1\linewidth]{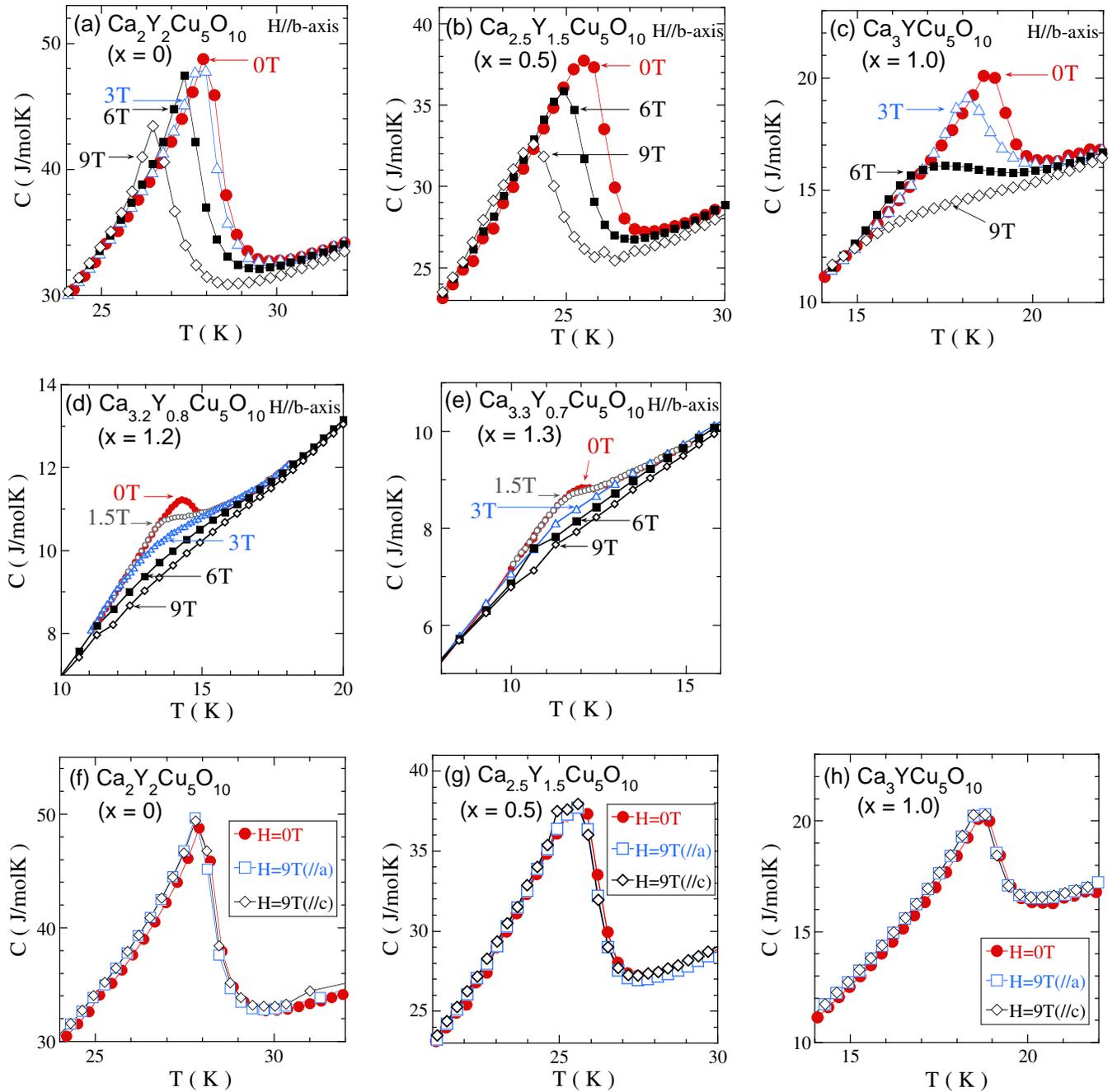}
\caption{(Color online) Temperature dependence of the specific heat, $C$, of Ca$_{2+x}$Y$_{2-x}$Cu$_5$O$_{10}$ with $0 \le x \le 1.3$ in magnetic fields up to 9 T (a)-(e) parallel to the $b$-axis and (f)-(h) parallel to the $a$- and $c$-axes.}
\label{fig:9}
\end{center}
\end{figure*}
\subsubsection{$x =$ 0, 0.5, 1.0, 1.2 and 1.3}
\begin{figure*}[p]
\begin{center}
\includegraphics[width=0.7\linewidth]{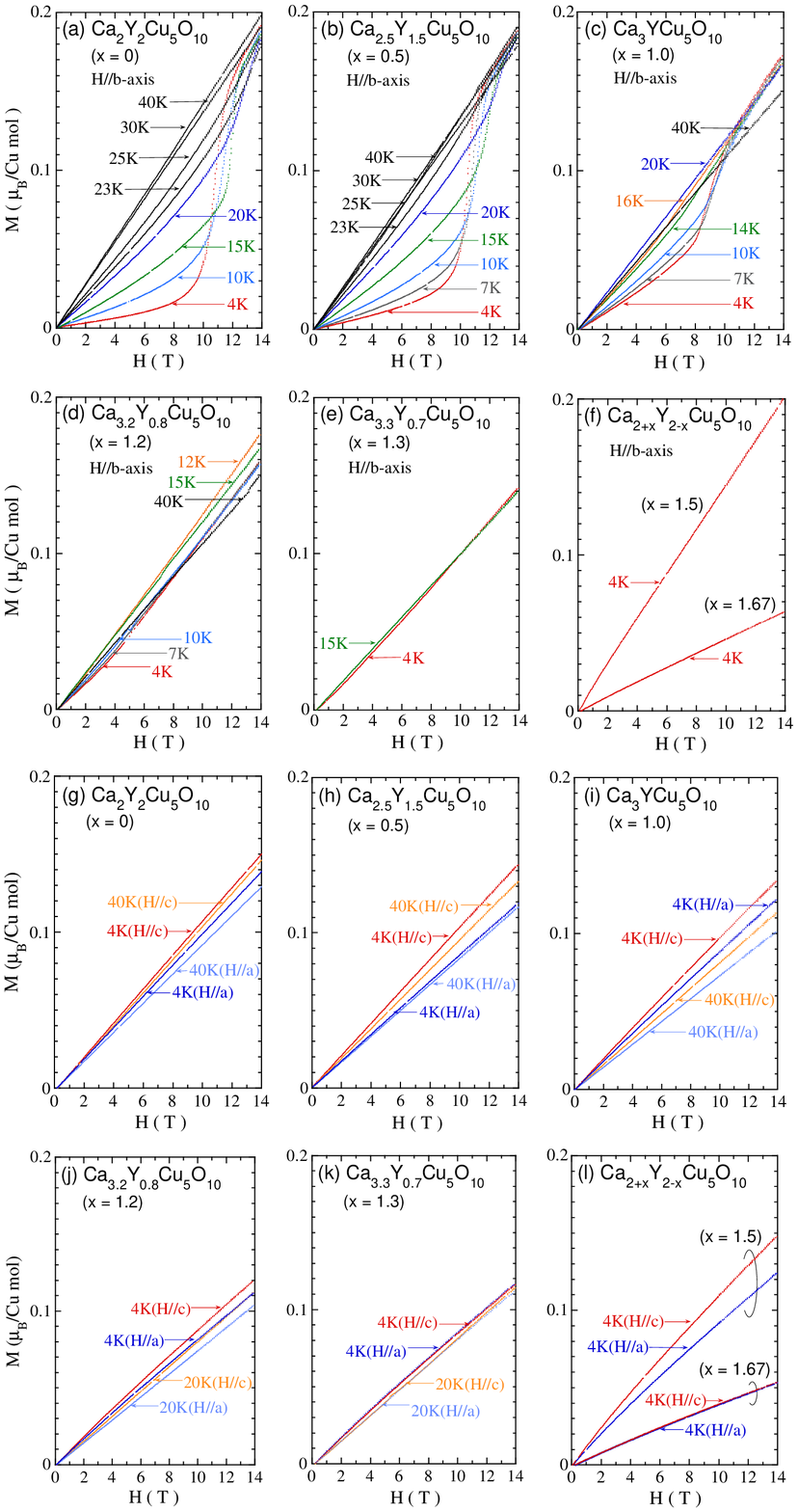}
\caption{(Color online) Magnetic-field dependence of the magnetization curve of  Ca$_{2+x}$Y$_{2-x}$Cu$_5$O$_{10}$ with $0 \le x \le 1.67$ at various temperatures in magnetic fields up to 14 T (a)-(f) parallel to the $b$-axis and (g)-(l) parallel to the $a$- and $c$-axes.
}
\label{fig:10}
\end{center}
\end{figure*}
Figures \ref{fig:9}(a)-(e) show the temperature dependence of the specific heat in magnetic fields parallel to the $b$-axis for $x =$ 0, 0.5, 1.0, 1.2 and 1.3. 
With increasing magnetic-field, the $\lambda$-type peak is reduced and shifted to lower temperatures. 
By the application of magnetic fields parallel to the $a$- and $c$-axes, on the other hand, the shift of the peak is very small, as shown in Figs. \ref{fig:9}(f)-(h). 
The anisotropic effect of magnetic field on the peak is consistent with the result that the magnetic easy axis is parallel to the $b$-axis. 
That is, in magnetic fields parallel to the $b$-axis, $T_{\rm N}$ is markedly lowered due to the competition between the exchange energy and the Zeeman energy. 
In magnetic fields parallel to the $a$- and $c$-axes (i.e. perpendicular to the spin direction), the component parallel to the $b$-axis of magnetic moments keeps the antiferromagnetic arrangement, though the magnetic moments gradually tend toward the field direction.
Accordingly, $T_{\rm N}$ does not decrease with increasing field so much.

Figures \ref{fig:10}(a)-(e) show the magnetization curves in magnetic fields up to 14 T parallel to the $b$-axis.
In $x =$ 0, 0.5, 1.0, a jump is clearly observed at low temperatures below $T_{\rm N}$\cite{thermal}. 
This magnetization jump is regarded as being due to a spin-flop transition, because the extrapolated line of the magnetization curve at high magnetic fields above the spin-flop transition field, $H_{\rm SF}$, tends to cross the origin. 
With increasing $x$, the spin-flop transition becomes smeared. 
In $x =$ 1.2 and 1.3, the spin-flop transition is not detected as a jump but a bend. 
By the application of magnetic fields parallel to the $a$- and $c$-axes, on the other hand, the magnetization increases linearly with increasing field, as shown in Figs. \ref{fig:10}(g)-(k). 
This is due to the gradual tendency of magnetic moments toward the field direction. 
Here, it is noted that Chabot {\it et al.}\cite{Chabot} have previously estimated $H_{\rm SF}$ as $\sim$ 5 T around 25--30 K in $x = 0$ from the magnetization curve of grain-aligned polycrystalline samples. 
However, such a transition is not detected in our measurements using a single crystal with $x = 0$ of good quality.

\begin{figure}[t]
\begin{center}
\includegraphics[width=1\linewidth]{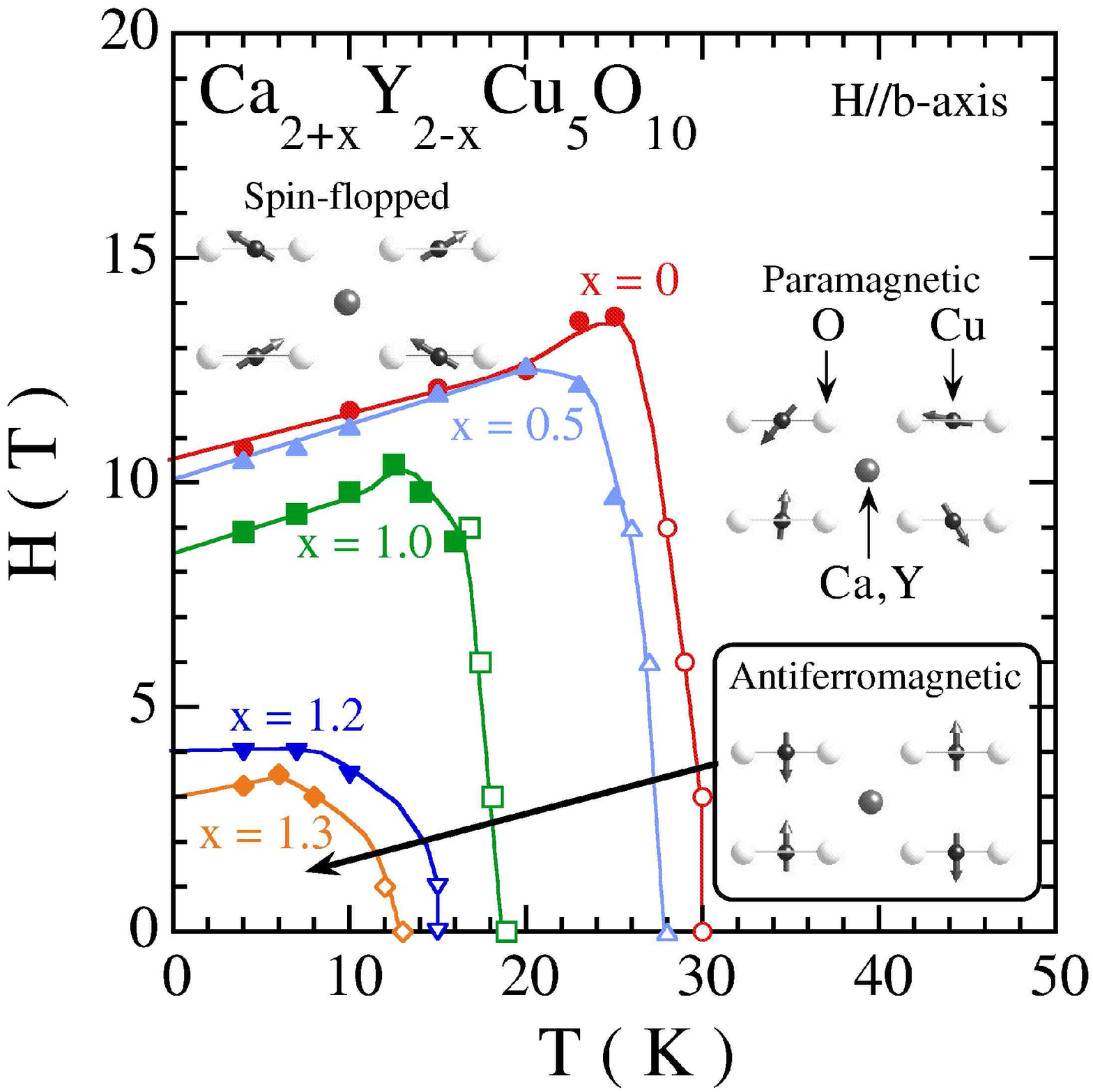}
\caption{(Color online) Magnetic phase diagram of  Ca$_{2+x}$Y$_{2-x}$Cu$_5$O$_{10}$ with $0 \le x \le 1.3$ as a function of magnetic field and temperature. 
Magnetic fields are applied parallel to the $b$-axis. 
Closed and open symbols indicate characteristic magnetic-fields and temperatures estimated from the magnetization curve and specific heat measurements, respectively. 
Typical spin arrangements seen along the $a$-axis (i.e. parallel to the CuO$_2$ chain) are shown in the paramagnetic, antiferromagnetically ordered and spin-flopped phases. 
}
\label{fig:11}
\end{center}
\end{figure}

Figure \ref{fig:11} summarizes the characteristic fields and temperatures in the antiferromagnetically ordered state. 
$H_{\rm SF}$'s shown by closed symbols are defined at the maximum point of the field-derivative of the magnetization curve for $0 \le x \le 1.0$ and at the point where linear extrapolations of the segments of the magnetization curve below and above $H_{\rm SF}$ intersect with each other for $x = 1.2$ and 1.3. 
Transition temperatures defined at the peak of the temperature dependence of the specific heat are plotted by open symbols. 
It is found that the initial spin arrangement changes into a spin-flopped one at $H_{\rm SF}$ with increasing field and that the region of the initial spin arrangement becomes narrow through the hole doping.

Here, we estimate the magnitude of the the exchange field and the magnetic anisotropy field parallel to the $b$-axis. 
The spin-flop transition takes place at $H_{\rm SF}$ where the gain of the Zeeman energy overcomes the loss due to the magnetic anisotropy energy responsible for the preferred spin direction. 
In terms of a simple uniaxial mean-field model\cite{uniaxial}, the spin-flop transition field at 0 K is given by $H_{\rm SF}$(0) $=$ (2$H_{\rm E}H_{\rm A}$ $-$ $H_{\rm A}^2$)$^{1/2}$, where $H_{\rm A}$ and $H_{\rm E}$ are the magnetic anisotropy field and the exchange field, respectively.
$H_{\rm E}$ is roughly given by $(1/2)H_{\rm 0}$, where $H_{\rm 0}$ is a field where the Cu$^{2+}$ moment is saturated. 
By extrapolating the magnetization curve to the saturated value estimated from the number of Cu$^{2+}$ spins, $H_{0}$ is estimated as $\sim$ 70 T for $x =$ 0, 0.5, 1.0, 1.2 and 1.3, which is almost independent of $x$. 
So, $H_{\rm E}$ is calculated to be $\sim$ 35 T, implying no change of the magnitude of the exchange interaction up to $x = 1.3$. 
Using the extrapolated value of $H_{\rm SF}$(0) $=$ 10.5 T, 10.0 T, 8.5 T, 4.0 T and 3.0 T, $H_{\rm A}$'s are calculated to be 1.6 T, 1.5 T, 1.0 T, 0.2 T and 0.1 T for $x =$ 0, 0.5, 1.0, 1.2 and 1.3, respectively. 
Therefore, it is suggested that the decrease of $T_{\rm N}$ through the hole doping is not attributed to the decrease of $J_1$ corresponding to $H_{\rm E}$ but the decrease of the magnetic anisotropy parallel to the $b$-axis corresponding to $H_{\rm A}$. 
This is consistent with the recent inelastic neutron scattering results of Ca$_{2+x}$Y$_{2-x}$Cu$_5$O$_{10}$ that $J_1$ is almost independent of $x$ and that the anisotropic exchange interaction, $D$, decreases with increasing $x$\cite{Matsudasingle}.

Meanwhile, a metamagnetic transition has been reported in the edge-sharing CuO$_2$ chain system of Ca$_9$La$_5$Cu$_{24}$O$_{41}$\cite{Am}. 
The difference in the transition in magnetic fields between Ca$_9$La$_5$Cu$_{24}$O$_{41}$ and Ca$_2$Y$_2$Cu$_5$O$_{10}$ is explained as being due to that in $J_1$.
That is, $J_1$ estimated from the neutron scattering experiments is as small as $D$ in Ca$_9$La$_5$Cu$_{24}$O$_{41}$\cite{CaLa5}, while $J_1$ is about 10-100 times larger than $D$ in Ca$_2$Y$_2$Cu$_5$O$_{10}$\cite{Matsudasingle}. 
Therefore, $H_{\rm E}$ and $H_{\rm A}$ are comparable with each other in Ca$_9$La$_5$Cu$_{24}$O$_{41}$, so that the metamagnetic transition takes place rather than the spin-flop transition. 
These results remind us that the magnetic interaction in the edge-sharing CuO$_2$ chain exhibits a remarkable change as a function of $\theta$.
%The difference in the transition in magnetic fields between Ca$_9$La$_5$Cu$_{24}$O$_{41}$ and Ca$_2$Y$_2$Cu$_5$O$_{10}$ may be explained as being due to the exchange interaction $J_{\rm b}$ along the $b$-axis, namely, along the magnetic easy axis. 
%That is, because of the existence of a two-leg spin-ladder plane between CuO$_2$ chain planes in Ca$_9$La$_5$Cu$_{24}$O$_{41}$, $J_{\rm b}$ in Ca$_9$La$_5$Cu$_{24}$O$_{41}$ is speculated to be smaller than $J_{\rm b}$ in Ca$_2$Y$_2$Cu$_5$O$_{10}$. 
%In fact, $T_{\rm N}$ in Ca$_9$La$_5$Cu$_{24}$O$_{41}$ is 9.7--10.5 K\cite{Am,MatsudaLa5} and lower than $T_{\rm N}$ in Ca$_2$Y$_2$Cu$_5$O$_{10}$. 
%Accordingly, it is likely that the antiferromagnetic coupling along the $b$-axis is more easily broken in Ca$_9$La$_5$Cu$_{24}$O$_{41}$ than in Ca$_{2}$Y$_{2}$Cu$_5$O$_{10}$, so that the metamagnetic transition takes place in the former rather than the spin-flop transition. 

\subsubsection{$x =$ 1.5 and 1.6}
\begin{figure}[b]
\begin{center}
\includegraphics[width=0.9\linewidth]{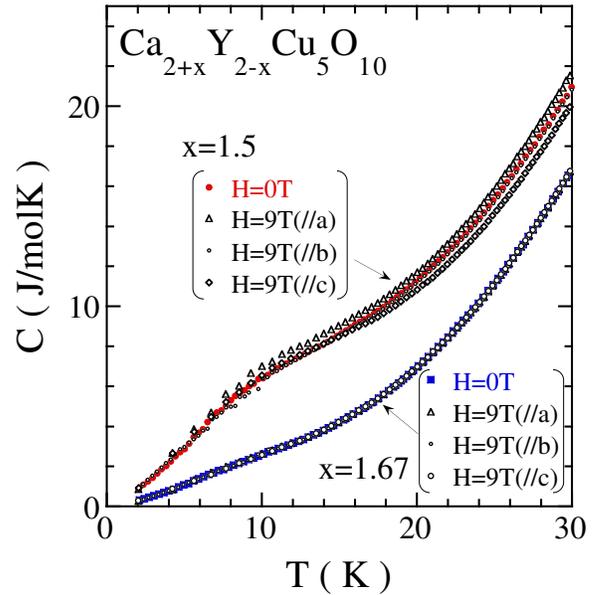}
\caption{(Color online) Temperature dependence of the specific heat, $C$, of  Ca$_{2+x}$Y$_{2-x}$Cu$_5$O$_{10}$ with $x =$ 1.5 and 1.67 in zero field and in a magnetic field of 9 T parallel to the $a$-, $b$- and $c$-axes. 
}
\label{fig:12}
\end{center}
\end{figure}
Figure \ref{fig:12} shows the temperature  dependence of the specific heat of  Ca$_{2+x}$Y$_{2-x}$Cu$_5$O$_{10}$ with $x =$ 1.5 and 1.67 in zero field and in a magnetic field of 9 T parallel to the $a$-, $b$- and $c$-axes. 
The broad peak observed around 10 K in zero field suggests the formation of a spin-glass state, as discussed in Sec. II B. 
The peak does not change by the application of a magnetic field of 9 T parallel to the $a$-, $b$- and $c$-axes. 
This may be reasonable, because the magnetic field of 9 T is much smaller than $|J_1|/\mu_{\rm B}$ which brings about short-range magnetic order. 
For $x = 1.5$ and 1.67, in fact, $|J_1|/k_{\rm B}$ has been estimated as 80 K\cite{Matsudasingle}. 
It is found that the spin-gap state in $x = 1.67$ is also not affected by the application of magnetic field. 
This is also reasonable, because the spin gap, $\Delta/\mu_{\rm B}$, is expected to be much larger than 9 T. 
In fact, it has been found that $\Delta/k_{\rm B}$ in the similar edge-sharing CuO$_2$ chain of Sr$_{14}$Cu$_{24}$O$_{41}$ is as large as $\sim$ 100 K\cite{KumagaiPRL,TsujiJLTP,Matsuda14n,Eccpoly,Matsuda14x}. 
Therefore, a possible peak of the specific heat due to the spin-gap formation in $x = 1.67$ may be located at a much higher temperature than the observed broad peak and hidden by the large contribution of $C_{\rm phonon}$.

Figures \ref{fig:10}(f) and (l) show the magnetization curve of Ca$_{2+x}$Y$_{2-x}$Cu$_5$O$_{10}$ with $x =$ 1.5 and 1.67 at 4 K in magnetic fields up to 14 T parallel to the $b$-, $a$- and $c$-axes. 
The magnetic-field dependence looks like a paramagnetic one. 
The paramagnetic behavior in $x = 1.5$ may be due to the weak interaction between short-range magnetically ordered regions in the spin-glass state, because the magnetization in $x = 1.5$ is much larger than that in $x = 1.67$ where the spin-glass phase is a minor one. 
As for $x = 1.67$, on the other hand, the strong suppression of the magnetic susceptibility suggestive of the spin-gap formation is observed as shown in Fig. \ref{fig:4} and discussed above. 
The suppression is also observed in the magnetization curve. 
However, the magnetization curve does not exhibit any kink up to 14 T, suggestive of the close of the spin gap, but the paramagnetic behavior due to a minor phase of the spin-glass.
This is reasonable, because $\Delta/\mu_{\rm B}$ is speculated to be much larger than 14 T as discussed from the specific heat measurements.

\section{Summary}
We have succeeded in growing large-size single-crystals of Ca$_{2+x}$Y$_{2-x}$Cu$_5$O$_{10}$ with $0 \le x \le 1.67$ by the TSFZ method and measured the magnetic susceptibility, specific heat and magnetization curve. 
In $0 \le x \le 1.3$, an antiferromagnetic phase transition with the magnetic easy axis along the $b$-axis has been detected in both magnetic susceptibility and specific heat measurements and a spin-flop transition has been observed in the magnetization curve. 
It has been found that $T_{\rm N}$ decreases with increasing $x$ and disappears around $x = 1.4$. 
Alternatively, a spin-glass phase appears around $x = 1.5$. 
The spin-glass state in $x = 1.3$ and 1.67 may be due to a minor phase.  
At $x = 1.67$ where the hole concentration is $\sim$ 1/3 per Cu, the pretty small magnetic susceptibility has indicated that a spin-gap state due to the formation of spin-singlet pairs seems to emerge as a main phase. 
No sign of the coexistence of an antiferromagnetically ordered state and a spin-gap state suggested in Ca$_{1-x}$CuO$_2$ has been found in Ca$_{2+x}$Y$_{2-x}$Cu$_5$O$_{10}$. 
This may be due to the intense modulation in the $ac$-plane and/or the randomness of Ca$^{2+}$ and Y$^{3+}$ ions in Ca$_{2+x}$Y$_{2-x}$Cu$_5$O$_{10}$.

 \acknowledgements
We are grateful to Professor K. Kumagai, Professor F. Matsubara, Professor H. Kato, Dr. M. Matsuda and Dr. Y. Miyazaki for the useful discussion.
We are indebted to Professor T. Nojima and Dr. S. Nakamura for the use of a vibrating sample magnetometer at Center for Low Temperature Science, Tohoku University. 
This work was supported by a Grant-in Aid for Scientific Research from the Ministry of Education, Culture, Sports, Science and Technology, Japan, and also by CREST of Japan Science and Technology Corporation.

%\bibliography{}

\begin{references}
\bibitem{stop} In the TSFZ growth of a layered compound, the cross section of the single-crystal rod tends to be ellipsoidal in shape. 
In this case, the rotation of the ellipsoidal rod is often hard to keep the molten zone stable. 
Empirically, the rotation had better be stopped in order to grow a large-size single-crystal.
\bibitem{residual} In fact, the inhomogeneity of $x$ induces residual Cu$^{2+}$-spins, which produce a minor phase of spin glass. 
At $x = 1.5$, on the other hand, a spin-glass phase emerges as a major phase because of a large number of residual Cu$^{2+}$-spins.
\bibitem{thermal} In $x = 0$, a jump has also been observed in the thermal conductivity at the same magnetic field as in the magnetization curve\cite{Kudo}.
\end{references}

\end{document}